# Optimization of noncollinear magnetic ordering temperature in Y-type hexaferrite by machine learning

Yonghong Li,[1] Jing Zhang,[1] Linfeng Jiang,[2] Long Zhang,[1] Yugang Zhang,[1] Xueliang Wu,[1] Yisheng Chai,[1,a)] Xiaoyuan Zhou,[2,a)] and Zizhen Zhou[2,a)]

[1]*Low Temperature Physics Laboratory, College of Physics, Chongqing University, Chongqing 401331, China*

[2]*Center of Quantum Materials and Devices, Chongqing University, Chongqing 401331, China.*

[a)]Authors to whom correspondence should be addressed: yschai@cqu.edu.cn, xiaoyuan2013@cqu.edu.cn and zzzhou@cqu.edu.cn

**Abstract:**

Searching the optimal doping compositions of the Y-type hexaferrite $Ba_2Mg_2Fe_{12}O_{22}$ remains a long-standing challenge for enhanced non-collinear magnetic transition temperature ($T_{NC}$). Instead of the conventional trial-and-error approach, the composition-property descriptor is established via a data driven machine learning method named SISSO (sure independence screening and sparsifying operator). Based on the chosen efficient and physically interpretable descriptor, a series of Y-type hexaferrite compositions are predicted to hold high $T_{NC}$, among which the $BaSrMg_{0.28}Co_{1.72}Fe_{10}Al_2O_{22}$ is then experimentally validated. Test results indicate that, under appropriate external magnetic field conditions, the $T_{NC}$ of this composition reaches up to reaches up to 568 K, and its magnetic transition temperature is also elevated to 735 K. This work offers a machine learning-based route to develop room temperature single phase multiferroics for device applications.

Multiferroicity and the related magnetoelectric (ME) effects have gained significant interest in materials science. Magnetoelectric multiferroics, which exhibit coexistence of ferroelectric and magnetic orders and show strong ME effects, are particularly intriguing. The ME cross-coupling between these orders in such materials offers considerable promise for novel functional devices.[1,2] Among the various types of magnetoelectric multiferroics, hexaferrites are prominent, especially due to their robust magnetoelectric effects at temperatures up to room temperature.[3-7]

Hexaferrites are categorized into six types: M, W, Y, Z, X, and U.[8,9] The Y-type hexagonal ferrite, represented by the chemical formula $Ba_2Me_2Fe_{12}O_{22}$ (where $Me$ can be $Co^{2+}$, $Zn^{2+}$, $Ni^{2+}$, etc.),[9] is the focus of extensive research. Several compositions with room temperature ME effects by non-collinear spin configurations are found previously.[10-12] Figure 1(a) illustrates its crystal and magnetic structure. Its crystal symmetry is characterized by the $R\bar{3}m$ space group with 6 different sites for Fe and $Me$ ions ($3b_{VI}$, $6c_{IV}$*, $6c_{VI}$, $18h_{VI}$, $6c_{IV}$, and $3a_{IV}$ sites). Its magnetic structure consists of alternating large ($L$) and small ($S$) spin blocks along the $c$-axis. In these blocks, spins ($Fe^{3+}$ and $Me^{2+}$) are arranged antiferromagnetically, leading to the formation of large magnetic moments $\mu_L$ in the $L$ blocks and small magnetic moments $\mu_S$ in the $S$ blocks.[13] This arrangement results in strong superexchange interactions and magnetic frustration at block boundaries, promoting various non-collinear magnetic structures at zero field. The confirmed non-collinear magnetic phases include proper screw, transverse conical, longitudinal conical, and alternative longitudinal conical phases.[14,15] Under an in-plane magnetic field, regardless of the initial phase, all these magnetic structures can transform into the transverse cone phases. Therefore, on one hand, the transverse cone phases can always persist up to $T_{NC}$ under a finite in-plane magnetic field. On the other hand, these transverse cone phases can host an in-plane polarization ($P$) via the inverse Dzyaloshinskii-Moriya (DM) interaction[16] or spin-current mechanism,[17] expressed as: $P \sim \mathrm{A} \sum_{ij} k \times (\mu_L \times \mu_S)$ ($k$: propagation vector). The in-plane magnetic field can easily tune the $P$ vector or its amplitude in these phases, leading to a strong ME effect up to $T_{NC}$.

However, the ME effects in most multiferroic Y-type hexaferrites manifest at low temperatures and under high magnetic fields due to low $T_{NC}$, limiting their practical applications. Addressing this challenge, research has shown that doping with non-magnetic and magnetic ions can be very effective to tune the $T_{NC}$ and required magnetic field. Kimura *et al.* first observed magnetoelectric coupling in $Zn_2Y$-type hexaferrite $Ba_{0.5}Sr_{1.5}Zn_2Fe_{12}O_{22}$, with a critical magnetic field for inducing polarization at low temperatures around 1 T.[5] Substituting Zn with Mg in $Ba_2Mg_2Fe_{12}O_{22}$, the Y-type hexaferrite, reduces this field to approximately 30 mT.[18,19] Chun *et al.* demonstrated the induction of a ferroelectric phase at 30 K with a mere 1 mT magnetic field by partially replacing Fe with Al in $Ba_{0.5}Sr_{1.5}Zn_2(Fe_{1-x}Al_x)_{12}O_{22}$, significantly enhancing the ME effect.[4] Further, replacing Zn with Co in the $BaSrZn_2Fe_{11}AlO_{22}$ sustains this effect up to room temperature. Meanwhile, its ME coefficient α (= d$P$/d$H$) reaches 400 ps/m at a 13 Oe magnetic field,[12] highlighting the impact of magnetic ion doping on the non-collinear magnetic structure and ME coupling. So far, the highest non-collinear magnetic structure is limited up to 430 K, which should be equal to $T_{NC}$. To further enhance $T_{NC}$, the present trial-and-error experimental approach nearly reaches its limitation.

Recently, a data-driven method named Sure Independence Screening and Sparsifying Operator (SISSO) has been widely used in functional materials discovery. With experimental or theoretical generated training data, it can assist the search of materials with ultralow thermal conductivity,[20] the prediction of the carrier relaxation time[21] and the stability of perovskite oxides and halides.[22] Therefore, adopting the above data-driven method to design and discovery new Y-type hexaferrites, is highly desirable.

In this letter, we performed a comprehensive search of literature concerning the doping effects of $Sr^{2+}$, $Ni^{2+}$, $Co^{2+}$, $Zn^{2+}$, $Al^{3+}$, $Cr^{3+}$, $Mn^{3+}$ and $Ni^{3+}$ on the parent compound. Leveraging advanced machine learning technique, SISSO, we determined an ideal composition $BaSrMg_{0.28}Co_{1.72}Fe_{10}Al_2O_{22}$ (BSMCFAO), that would achieve the highest non-collinear magnetic transition temperature. We observed that the magnetization curves of BSMCFAO polycrystal undergo a transition from a transverse conical structure to a longitudinal conical structure at around 362 K, and finally to a collinear ferrimagnetic structure with increasing temperature up to 568 K. Finally, BSMCFAO demonstrates a magnetic ordering temperature of up to 735 K. Our research points to a possible $T_{NC}$ as high as 568 K with suitable choice of external magnetic field.

To find the optimal chemical composition with the highest $T_{NC}$, we first conducted an exhaustive search of nearly all the experimentally reported BaMgFeO-based Y-type hexaferrites exhibiting non-collinear spin configurations. We then regard the non-collinear magnetic ordering as the transverse cone phase since they can always be transformed into a transverse cone spin configuration under certain in-plane magnetic field strength.[10,16,23] Table S1 shows all the chemical compositions and their $T_{NC}$ created by doping Sr in the Ba site, Ni/Zn/Co in the Mg site, and Al/Cr/Mn/Ni in the Fe site.

Then, a deeper statistical correlation analysis is applied to Table S1, as depicted in Fig. 1(b). The calculated Pearson's correlation coefficient reveals strong or weak correlations between the doping elements with the non-collinear transition temperature. A related physical analysis can also be conducted based on the literatures. For Ba/Sr site, it suggests that a balanced ratio of Ba and Sr, as observed in the $Ba_{2-x}Sr_xZn_2Fe_{12}O_{22}$ system, could achieve a high $T_{NC}$. Notably, the non-collinear spin configuration in this system is present only at intermediate values of $x$, not at the end members.[24] Regarding the Mg site, our analysis indicates that while Ni and Zn negatively affect $T_{NC}$, Mg and Co positively enhance it. Intriguingly, despite $Ni^{2+}$ being a magnetic ion, it adversely impacts the non-collinear spin configuration. Previous reports have shown that $Ni^{2+}$ can raise the Curie temperature when substituting non-magnetic $Zn^{2+}$ ion in $Ba_2Zn_2Fe_{12}O_{22}$,[25] thereby strengthening the superexchange interaction between magnetic blocks. To understand its anomalous role concerning $T_{NC}$, it is essential to consider $Ni^{2+}$'s preferential occupation of octahedral sites (mainly $3a_{VI}$, $18h_{VI}$, and $3b_{VI}$).[26,27] While $Ni^{2+}$ does not directly modify the superexchange interaction between the $L$ and $S$ blocks, it influences their magnetic anisotropy. In contrast, $Co^{2+}$ primarily occupies the central octahedral position ($6c_{VI}$) at the boundary of the $L$ and $S$ blocks,[27] introducing strong magnetic anisotropy and significantly influencing the superexchange interaction, which is vital for the non-collinear spin configuration in Y-type hexaferrite.

The role of non-magnetic $Mg^{2+}$ is also noteworthy. Its pronounced preference for the octahedral ($18h_{VI}$) site inside the magnetic $L$ block likely reduces the in-plane magnetic anisotropic energy of this block. Conversely, non-magnetic $Zn^{2+}$ predominantly occupies the tetrahedral $6c_{IV}$ and $6c_{IV}$* sites within and at the boundary of the $L$ and $S$ blocks, respectively. This occupancy could alter the exchange interaction between these blocks. For $Al^{3+}$, our previous experimental studies indicate a dominant preference for the octahedral position at the $L$ and $S$ block boundary.[4] X-ray absorption spectroscopy at the Fe $L_{2,3}$ edge suggests that $Fe^{3+}$ at this site possesses a significant orbital moment, and Al substitution fine-tunes the magnetic anisotropy, thereby improving $T_{NC}$.

The above analysis prompts the hypothesis that although the underlying physical phenomena that give rise to $T_{NC}$ are complex, a physically motivated descriptor based on the chemical composition could be predictive in Y-type hexaferrite. Rely on 83 instances, we can then determine the optimal composition from the revealed descriptor and check it experimentally.

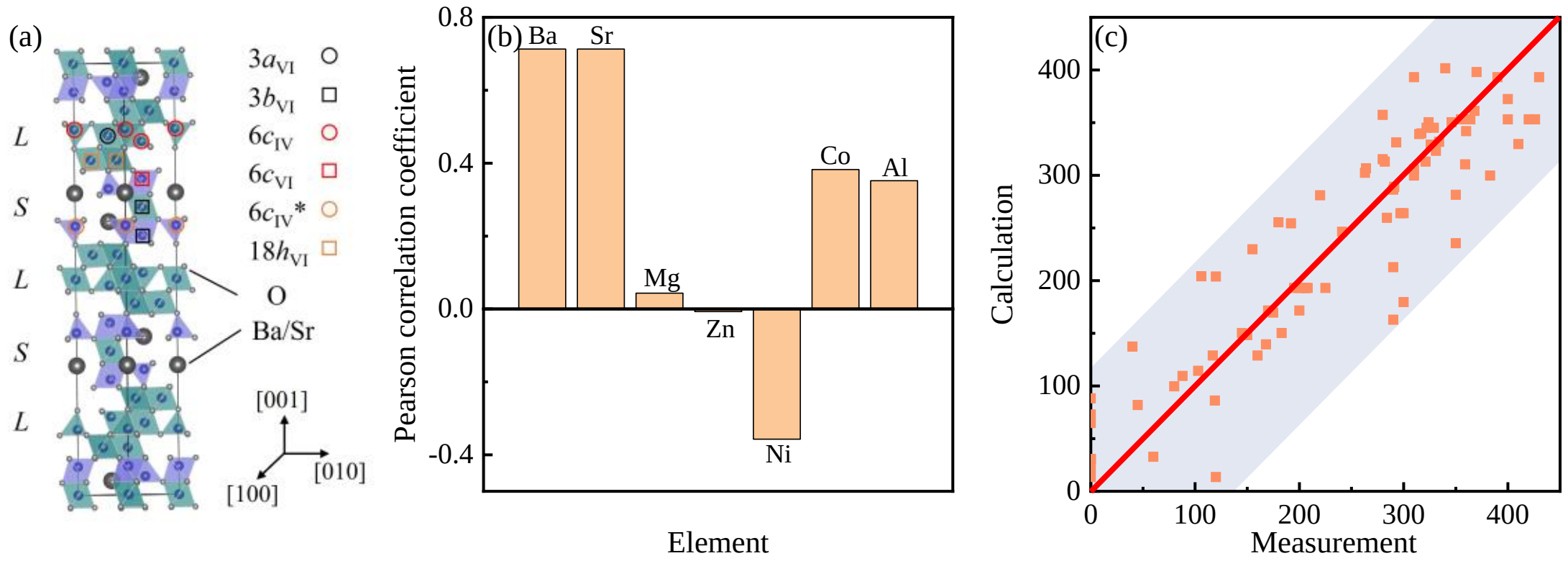

**FIG. 1.** (a) Schematic crystal structure of Y-type hexaferrite. (b) The correlation between experimentally measured $T_{NC}$ and the content of each element. (c) The intuitive linear correlation between the experimentally measured $T_{NC}$ and predicted $T_{NC}$.

Utilizing the data-driven machine learning approach SISSO, we comprehensively considered various atomic characteristic parameters, including atomic number, atomic mass, atomic orbital angular momentum, elemental content, and so on. Our findings indicate that focusing solely on elemental contents as the features yield the best model. We processed these training data defined by the content of each element, given their fixed elemental composition. Then, we input all the features to construct various descriptors. Initially, we developed over $10^{10}$ features with a complexity of three using the operators F = {+, –, ×, /, exp, log, | |, $^{0.5}$, $^{-1}$, $^{2}$, $^{3}$}. From these, we selected a series of descriptors with high Pearson correlation coefficients (detailed in the supplementary material). The subsequent phase of our research focuses on identifying the optimal descriptor, emphasizing its physical significance.

The SISSO training performed on the collected $T_{\mathrm{NC}}$ data of 83 BaMgFeO-based Y-type hexagonal ferrites led us to identify an optimal descriptor for $T_{\mathrm{NC}}$:

$$T_{\mathrm{NC}} = 70.715 \times \left[ \left| n_{Co} + n_f - |n_{Co} - n_{Sr}| \right| - \left| n_{Ba} + 2 - n_{Mg} - n_{Co} - n_{Sr} \right| \right] + 294.42 \quad (1)$$

Here, '$n$' represents the content, and the subscript denotes the corresponding element and '$f$' represents the elements replace the $Fe^{3+}$ sites. As illustrated in Fig. 1(c) and Table S1, the $T_{\mathrm{NC}}$ values calculated using this descriptor closely align with experimental measurements, yielding a Pearson correlation coefficient above 90%. Additionally, we calculated the mean absolute error and root mean square error, which are 37.136 and 45.586, respectively. These results corroborate the robust predictive capability of our descriptor.

Through Monte Carlo simulations (detailed in the supplementary material), $BaSrMg_{0.28}Co_{1.72}Fe_{10}Al_2O_{22}$ (BSMCFAO) is emerged as the composition with the highest predicted non-collinear transition temperature ($T_{\mathrm{NC}}$), approximately 507 K. This prediction is based on the limitation of $n_f$ (the number of $Fe^{3+}$ ions fully replaced by Al at the $6c_{\mathrm{VI}}$ site) to a maximum of two. From our predictive model, we derived three key rules: 1) Mg and Co are the most effective 2+ ions for enhancing $T_{\mathrm{NC}}$, 2) a 1:1 ratio of Ba and Sr optimizes $T_{\mathrm{NC}}$, and 3) complete replacement of $Fe^{3+}$ at the $6c_{\mathrm{VI}}$ site with Al positively influences $T_{\mathrm{NC}}$. These rules align with the trends indicated by our Pearson correlation coefficients. It's important to note that our dataset lacks sufficient data to conclusively predict the effects of Cr/Mn/Ni doping at the Fe site, and thus these elements are not included in our equation. However, considering their potential valence states, these ions might induce oxygen vacancies, potentially enhancing the conductivity of the sample.

We synthesized polycrystalline BSMCFAO using a solid-state reaction and confirmed its composition via X-ray diffraction. Figure 2(a) displays the powder X-ray diffraction pattern of the sample at room temperature. The prime peak position of the main impurities during the growth process of Y-type hexaferrites—$BaFe_2O_4$ and α-$Fe_2O_3$, are marked as ◆ and ▼ respectively. It can be seen from the figure that there is no significant intensity of XRD at those positions. Additionally, we conducted Energy Dispersive Spectrometer (EDS) measurements on the samples to determine their elemental composition. The results indicate that the actual chemical stoichiometry is close to the theoretical one. Please refer to the supplementary materials for details. Figures 2(b) and (c) show the temperature ($T$) dependence of the field-cooled (FC) magnetization curves, measured under magnetic fields of 1000 Oe and 83 Oe, respectively. These measurements span a temperature range from 10 to 800 K, following a high field history. To prepare for these measurements, an external magnetic field of $H$ = 50 kOe was applied at the lowest temperature to remove the initial state, after which the field was reduced to 1000 Oe or 83 Oe.

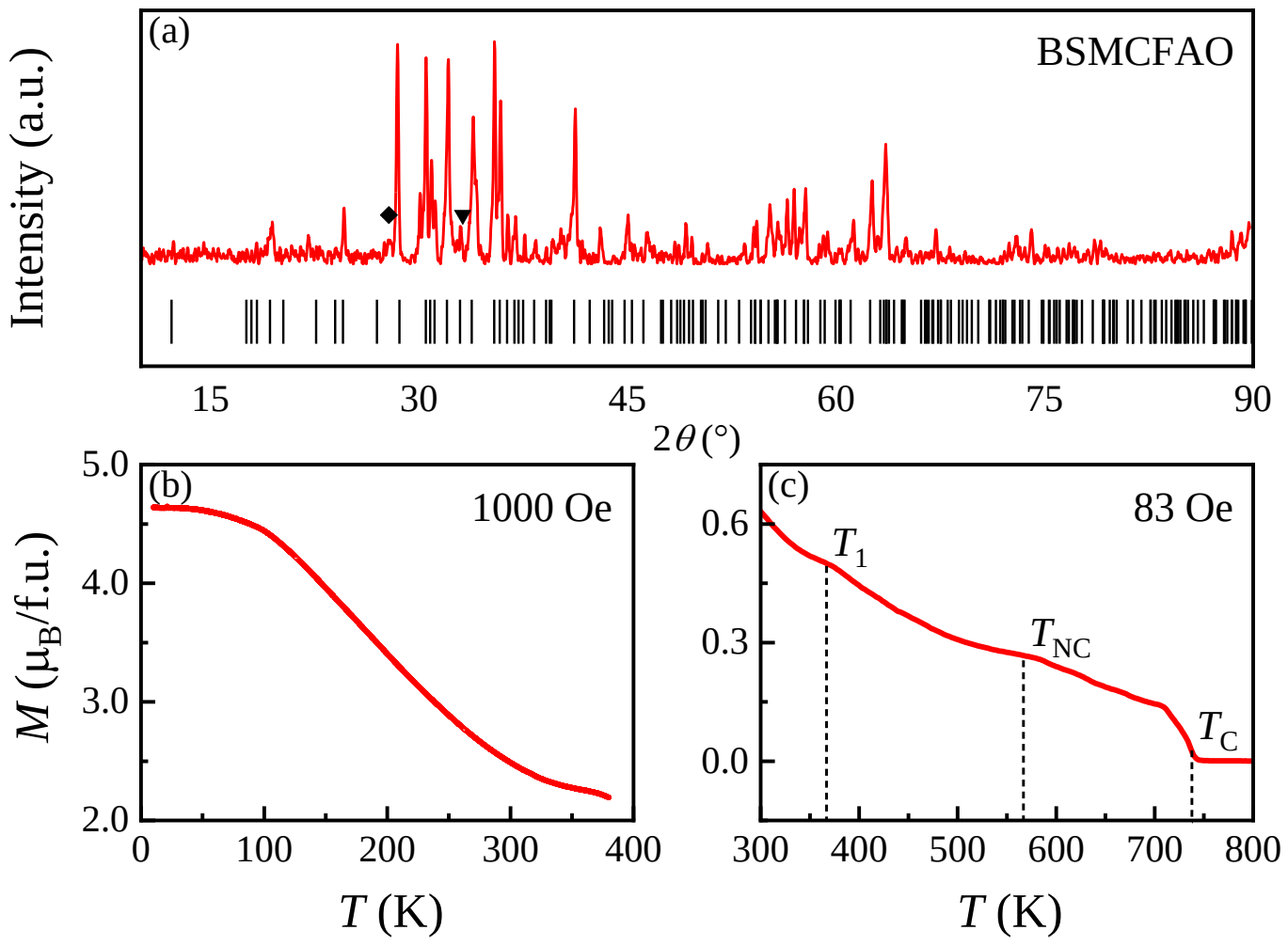

**FIG. 2.** (a) Powder X-ray diffraction pattern of BSMCFAO at room temperature. (b) and (c) Temperature dependence of the magnetization under external magnetic field of 1000 Oe and 83 Oe for BSMCFAO after a high field history, respectively.

The magnetization ($M$) as a function of temperature generally decreases with increasing temperature. Notably, two distinct anomalies are observed: the first around $T_1 \sim 362$ K, and the second around the $T_{NC} \sim 568$ K. These anomalies are followed by a sharp decline around 700 K, with the curve eventually flattening at the Curie temperature ($T_C$) of about 735 K. All these transition temperatures can be clearly found in the derivative of $M$ vs. $T$ curve in the Fig. S3. Previous research[28] has reported the magnetic phase diagram of Y-type hexaferrite $BaSrCo2Fe_{12-x-\delta}Al_xCr_\delta O_{22}$ with $x = 0.9$ and $\delta = 0.00$ and 0.05 with very similar composition to our $BaSrMg_{0.28}Co_{1.72}Fe_{10}Al_2O_{22}$. After field cooling in 100 Oe and in warming process, the sample will change from alternative longitudinal conical to proper screw, collinear ferrimagnetic and finally paramagnetic phase in series. In our case, we applied a stronger magnetic field and reduced to 1000 Oe or 83 Oe that a transverse cone or a transverse cone + alternative longitudinal cone will be brought down to this low field around room temperature.[29] Then, the sample undergoes a transition from a mixed conical state to a proper screw phase at 362 K and a transition from the proper screw phase to the ferrimagnetic phase at 568 K. This temperature is also close to the calculated value of $T_{NC}$. Beyond this point, the magnetization intensity diminishes rapidly, approaching zero at $T_C$.

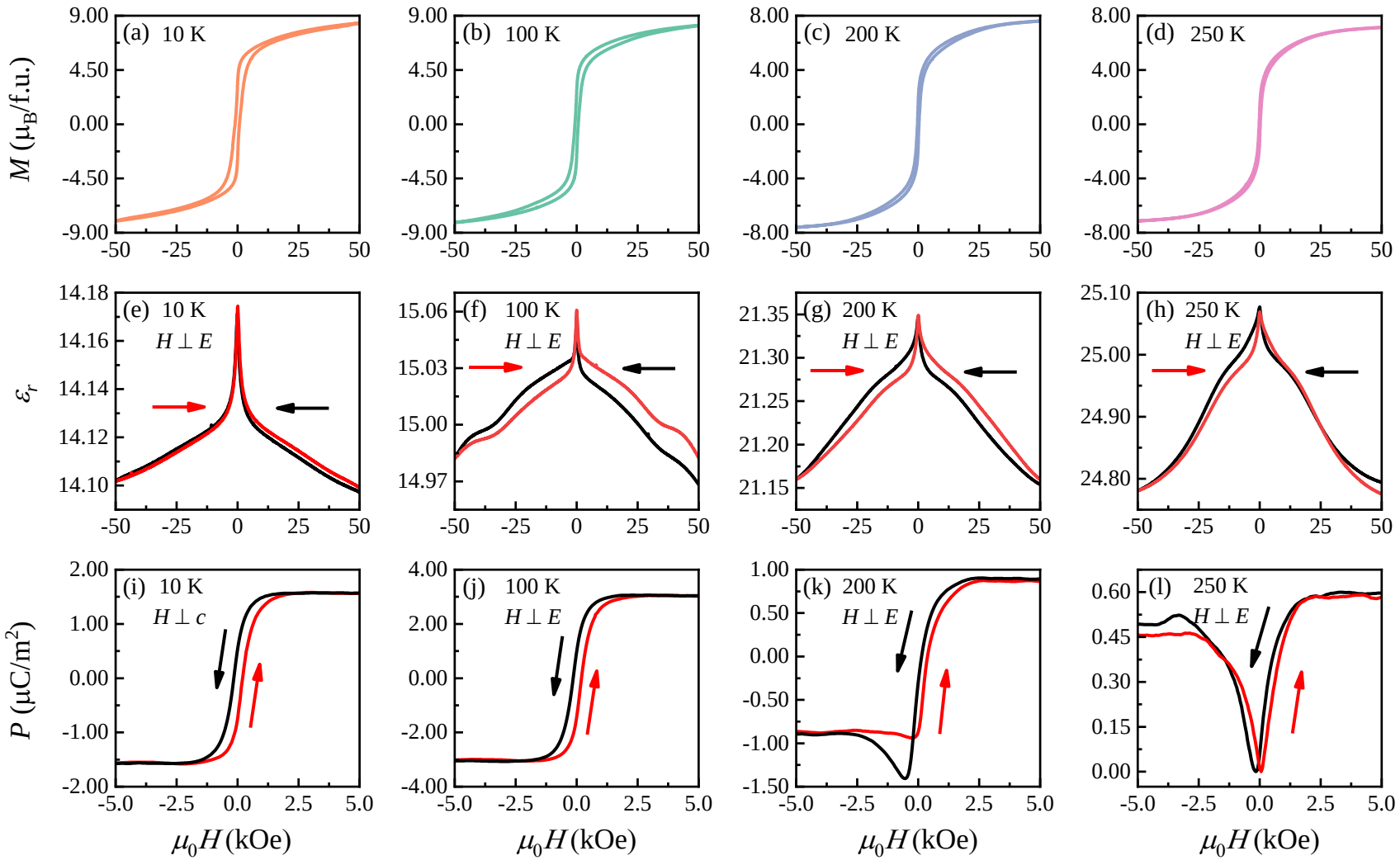

**FIG. 3.** (a)-(d) $H$ dependence of magnetization $M$ at 10 K, 100 K, 200 K and 250 K, respectively. (e)-(h) $H$ dependence of the relative dielectric constant $\varepsilon_r$ at selected temperatures. Arrows indicate the direction of swept magnetic field. (i)-(l) $H$ reversal of the electrical polarization $P$ at selected temperatures.

Figures 3(a)-(d) illustrate the dependence of magnetization ($M$) on the magnetic field ($H$) at temperatures of 10 K, 100 K, 200 K, and 250 K. At lower magnetic fields, the magnetization intensity shows a rapid increase, eventually approaching a saturation value ($M_S$) as the magnetic field strength increases. This behavior can be attributed to the transition from a transverse spin cone exhibiting ferroelectricity to a collinear ferrimagnetic state with paraelectricity. At 10 K and a magnetic field of $H_{ab}$ = 10 kOe, the cone angle, estimated at approximately 43°, aligns closely with that observed in $Ba_{0.5}Sr_{1.5}Zn_2(Fe_{0.92}Al_{0.08})_{12}O_{22}$, as indicated by the magnetization being 0.72 times $M_S$.[30]

Figures 3(e)-(h) present the magnetic field dependence of the relative dielectric constant ($\varepsilon_r$) at the same set of temperatures. The high temperature dielectric data up to 380 K is available in the supplementary material. A sharp dielectric peak near zero field is observable at all temperatures, which gradually diminishes with increasing magnetic field strength. The dielectric variations in BSMCFAO correspond with the changes in magnetization under varying magnetic fields. Notably, the sharp dielectric peak and dramatic magnetization reversal around zero field typically signify the presence of transverse cone-induced ferroelectricity at low magnetic fields. Therefore, a strong magnetodielectric effect up to 380 K corroborates the non-collinear spin configuration above $T_1$ ~ 362 K and very like up to $T_{NC}$ ~ 568 K.

To further explore the field dependent polarization ($P$) near zero field at selected temperatures, we measured the magnetoelectric current ($J_{ME}$) and integrated it over time to ascertain the $P(H)$ curves, as depicted in Figs. 3(i)-(l). Prior to these measurements, the sample was magnetoelectrically poled under an electric field of E = 43.5 kV/m. Due to the sample's less insulating nature, the poling field used was ten times smaller than that typically applied to other polycrystalline samples.[31] Above 250 K, reliable polarization to derive a $P$-$H$ curve was not achievable. Despite the reduced poling field, a maximum polarization of 3 μC/m² was observed at 100 K, significantly lower than expected for a single magnetoelectric domain but sufficient to demonstrate the sample's ferroelectricity. We have summarized the maximum magnetoelectric coefficients and the reported $T_{NC}$ of other BaMgFeO-type materials in Table S2. Although the magnetoelectric coefficient of our sample is small, it is mainly due to the much lower poling electric field (43.5 kV/m) compared to 750 kV/m in $Ba_{0.4}Sr_{1.6}Mg_2Fe_{12}O_{22}$.[29] Unfortunately, our sample is much leakier than other reported compounds in the Table S2. The measured magnetoelectric coefficients of polycrystalline samples are also much lower than those of single crystals. Still, if one can synthesize single crystals with the same chemical composition, it is very likely that the obtained magnetoelectric coefficient would be much higher. Moreover, below 200 K, polarization of this sample could be fully reversed at low magnetic fields, confirming the presence of a transverse cone phase at zero *H*. At 200 K, complete polarization reversal was not achieved, suggesting the partial existence of an alternative longitudinal cone phase or a proper-screw phase at zero *H*.[32] At 250 K, the polarization fully restored to its original value post-*H* reversal, indicating a dominant alternative longitudinal cone phase around zero *H*.[33]

Note that, if we fix the $n_{Sr} = n_{Ba} = 1$ and let $n_{Co} > 1$ in Eq. 1, it turns out to be much simplified:

$$T_{NC} = 70.715 \times \left[(n_f + 1) - (n_{Zn} + n_{Ni})\right] + 294.42. \qquad (2)$$

This Eq. 2 implies that a series of $BaSrMg_xCo_{2-x}Fe_{10}Al_2O_{22}$ ($x < 1$) will also show high $T_{NC}$. Also, more doping efforts should be tested on Fe sites other than Al since there is not enough data for us to distinguish the different role of Al/Cr/Mn/Ni at this site.

In conclusion, this study successfully utilized machine learning techniques to identify an optimal composition for Y-type hexaferrite, specifically $BaSrMg_{0.28}Co_{1.72}Fe_{10}Al_2O_{22}$, which exhibits a high $T_{NC}$ of 568 K. The successful application of machine learning in magnetoelectric multiferroics opens new avenues for enhancing the performance and capabilities of Y-type hexagonal ferrite-based devices.

## SUPPLEMENTARY MATERIAL

See the supplementary material for additional information about the SISSO training process, Monte Carlo simulation, crystal growth method, experimental detail, summary of extracted and the descriptor predicted values of $T_{NC}$ of BaMgFeO-based Y-type hexaferrites and summary of the maximum magnetoelectric coefficients α and the reported $T_{NC}$ of BaMgFeO-based Y-type hexaferrites in polycrystalline (Poly.) or single crystal (Single) form.

## ACKNOWLEDGMENTS

This work was supported by the Natural Science Foundation of China under Grants No. 11974065, No. 12227806, and No. 11874357. Y.S.C. acknowledges the support from Beijing National Laboratory for Condensed Matter Physics. We would like to thank G. W. Wang and Y. Liu at Analytical and Testing Center of Chongqing University for their assistance.

## AUTHOR DECLARATION

### Conflict of Interest

The authors have no conflicts to disclose.

### Author Contributions

**Yonghong Li**: Data curation (equal); Methodology (equal); Writing – original draft (equal); Writing – review & editing (equal). **Jing Zhang**: Data curation (equal). **Linfeng Jiang**: Software (equal). **Long Zhang**: Measurement (equal). **Yugang Zhang**: Measurement (equal). **Xueliang Wu**: Measurement (equal). **Yisheng Chai**: Conceptualization (equal); Supervision (equal); Writing – original draft (equal); Writing – review & editing (equal). **Xiaoyuan Zhou**: Conceptualization (equal); Supervision (equal). **Zizhen Zhou**: Software; Conceptualization (equal); Supervision (equal); Writing – original draft (equal); Writing – review & editing (equal).

## DATA AVAILABILITY

The data that support the findings of this study are available from the corresponding author upon reasonable request.

134433 (2019).

# Supplementary material

## Supplementary Note 1: SISSO training process

During the SISSO training, the intrinsically linear relationship observables in the compressed-sensing formalism was made nonlinear by equipping the features with nonlinear operators $H \equiv \{+, -, \times, /, \exp, \log, ||, ^{1/2}, ^{-1}, ^{2}, ^{3}\}$. At each iteration, $H$ operates on all available combinations, and more than $10^{10}$ features were constructed up to a complexity cutoff of 3. The sure independence screening (SIS) scores each feature with a correlation magnitude and keeps only the top ranked. The subset extracted by the SIS algorithm was set to 80,000. After the reduction, the sparsifying operators (SO) was used to pinpoint the optimal descriptor.

## Supplementary Note 2: Monte Carlo simulations

We randomly selected values for the elemental content and calculated $T_{NC}$ by substituting them into the formula. Subsequently, we randomly altered the elemental content and recalculated $T_{NC}$ to obtain a new value. By comparing the $T_{NC}$ values before and after these changes, we identified the elemental proportions leading to a larger $T_{NC}$. Iterating through these steps in a loop, we determined the elemental composition that results in the maximum $T_{NC}$, namely $BaSrMg_{0.28}Co_{1.72}Fe_{10}Al_2O_{22}$.

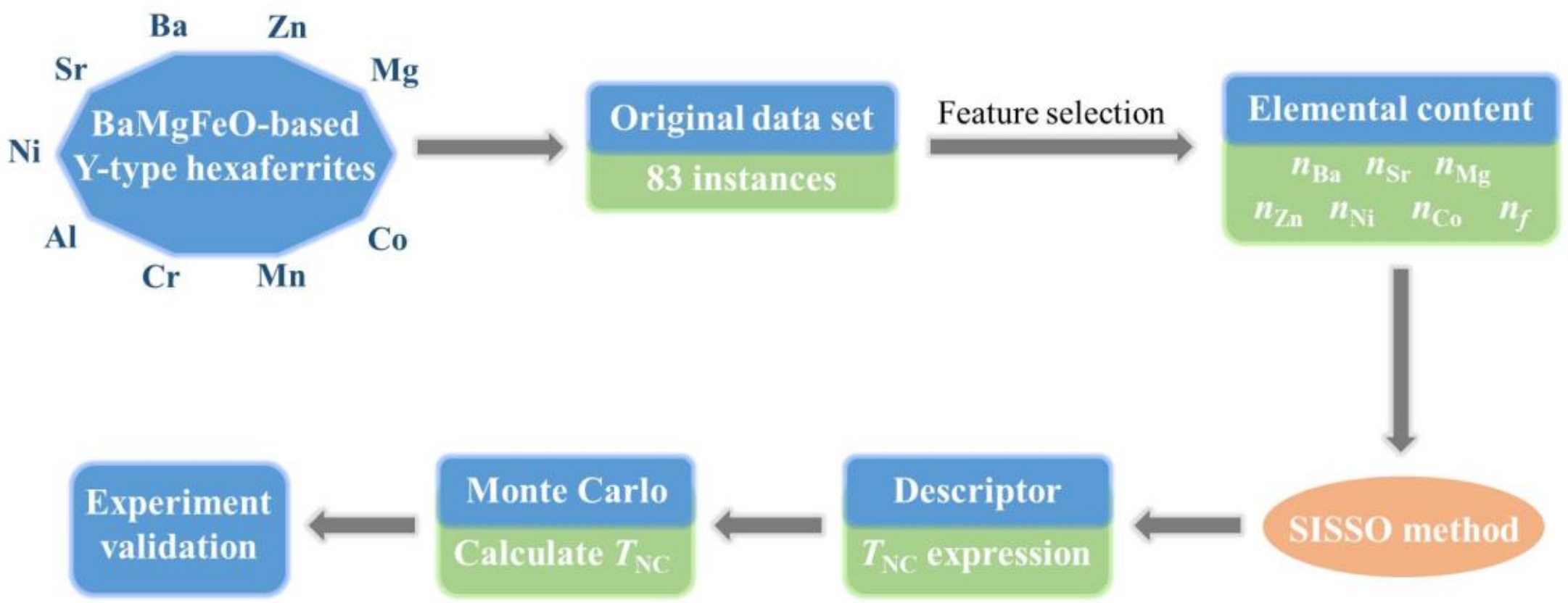

Figure S1. Flow chart of the SISSO.

## Supplementary Note 3: Sample growth method

Polycrystalline $BaSrMg_{0.28}Co_{1.72}Fe_{10}Al_2O_{22}$ were prepared using the solid-state reaction method. Chemically stoichiometric powders of $BaCO_3$, $SrCO_3$, MgO, $Co_2O_3$, $Al_2O_3$ and $Fe_2O_3$ were thoroughly mixed in an agate mortar and calcined at 1000 °C in air for 12 hours. The resulting mixture was re-ground after calcination, then pressed into cylindrical shapes and sintered at 1200 °C in air for 12 hours. To enhance resistivity, an annealing process was carried out at 900°C in an oxygen gas flow for 8 days, followed by slow cooling to room temperature at a rate of 50 °C/h.

**Supplementary Note 4: Experimental details**

Magnetization measurements (*M*-*T* and *M*-*H*) were conducted using a magnetic property measurement system (MPMS-3, Quantum Design). The $\varepsilon_r$ -*H* (relative dielectric constant) and *P*-*H* (ME current) were measured using an LCR meter (Agilent 4980A) and an electrometer (Keithley 6517B), respectively, within the Physical Property Measurement System (PPMS, Quantum Design). In order to measure $\varepsilon_r$-*H* (relative dielectric constant) and *P*-*H*, electrodes were attached to the largest surface of the sample using silver paste. During all measurements, the magnetic field is perpendicular to the *c*-axis.

**EDS:** To determine the elemental composition of the polycrystalline sample $BaSrMg_{0.28}Co_{1.72}Fe_{10}Al_2O_{22}$, we conducted EDS measurements, and the results are shown in Figure S2 as follows: O: 51.8%; Fe: 28.6%; Al: 7.0%; Co: 5.1%; Ba: 3.5%; Sr: 3.2%; Mg: 0.8%. These results indicate that the actual chemical stoichiometry is close to the theoretical one.

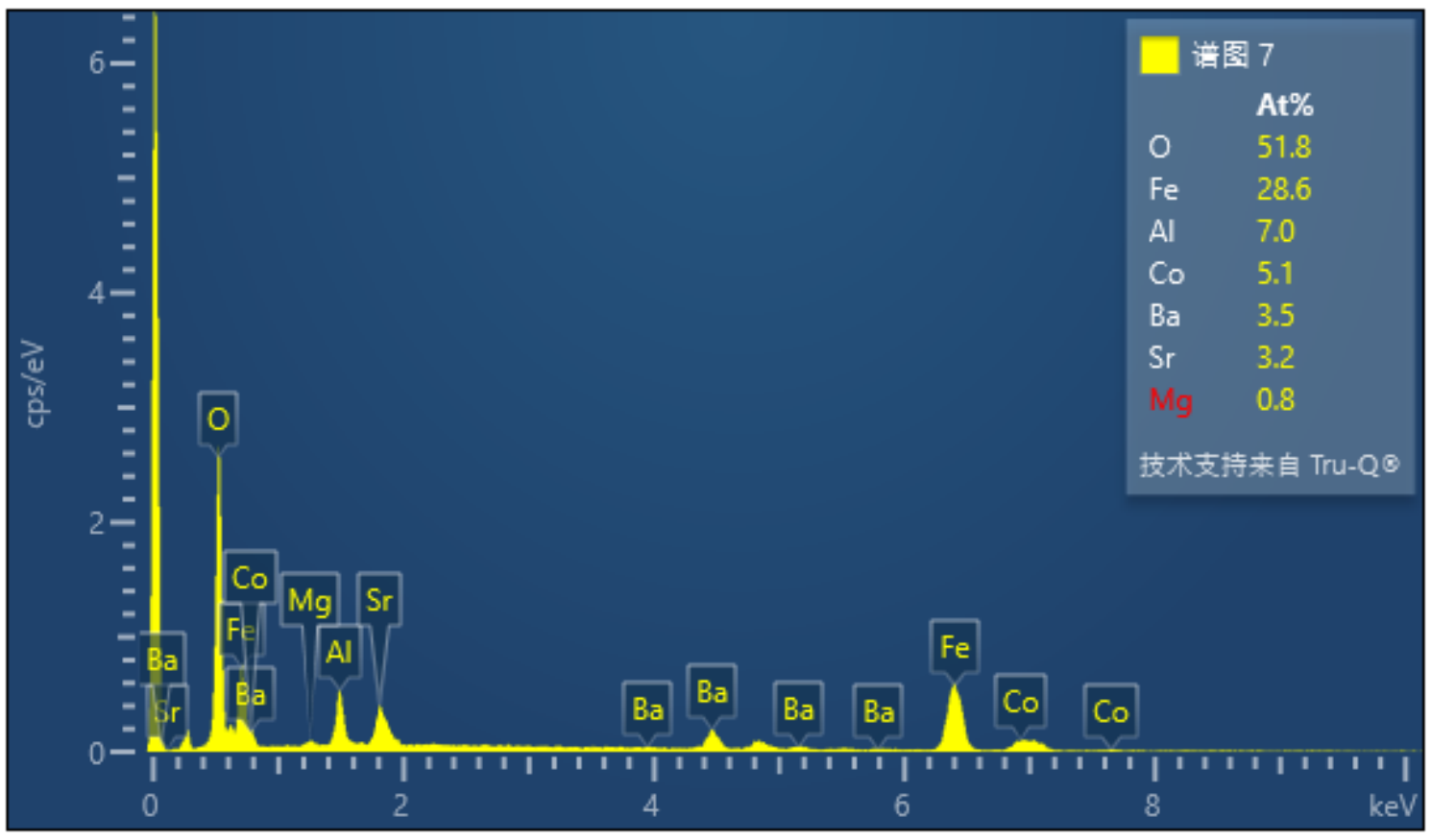

Figure S2. The EDS results for the $BaSrMg_{0.28}Co_{1.72}Fe_{10}Al_2O_{22}$ sample.

**Magnetization measurements:** During the *M-H* measurement, the complete magnetization loop is obtained by sweeping the field in the range of -50 kOe to 50 kOe. The *M*-*T* curve, on the other hand, is obtained by sweeping the temperature in the range of 10 K to 800 K. Prior to the *M*-*T* measurements, an external magnetic field of $H = 50$ kOe was applied at 10 K to induce the corresponding transverse cone state. The field was then gradually reduced to 1000 Oe or 83 Oe for the measurements. Following that, formal measurements were commenced.

Figure S3 displays the first derivation of *M* vs *T* curve. We determined the values of the three temperatures based on the changes in the slope of the curve, which are $T_1 = 362$ K, $T_{NC} = 568$ K, and $T_C = 735$ K.

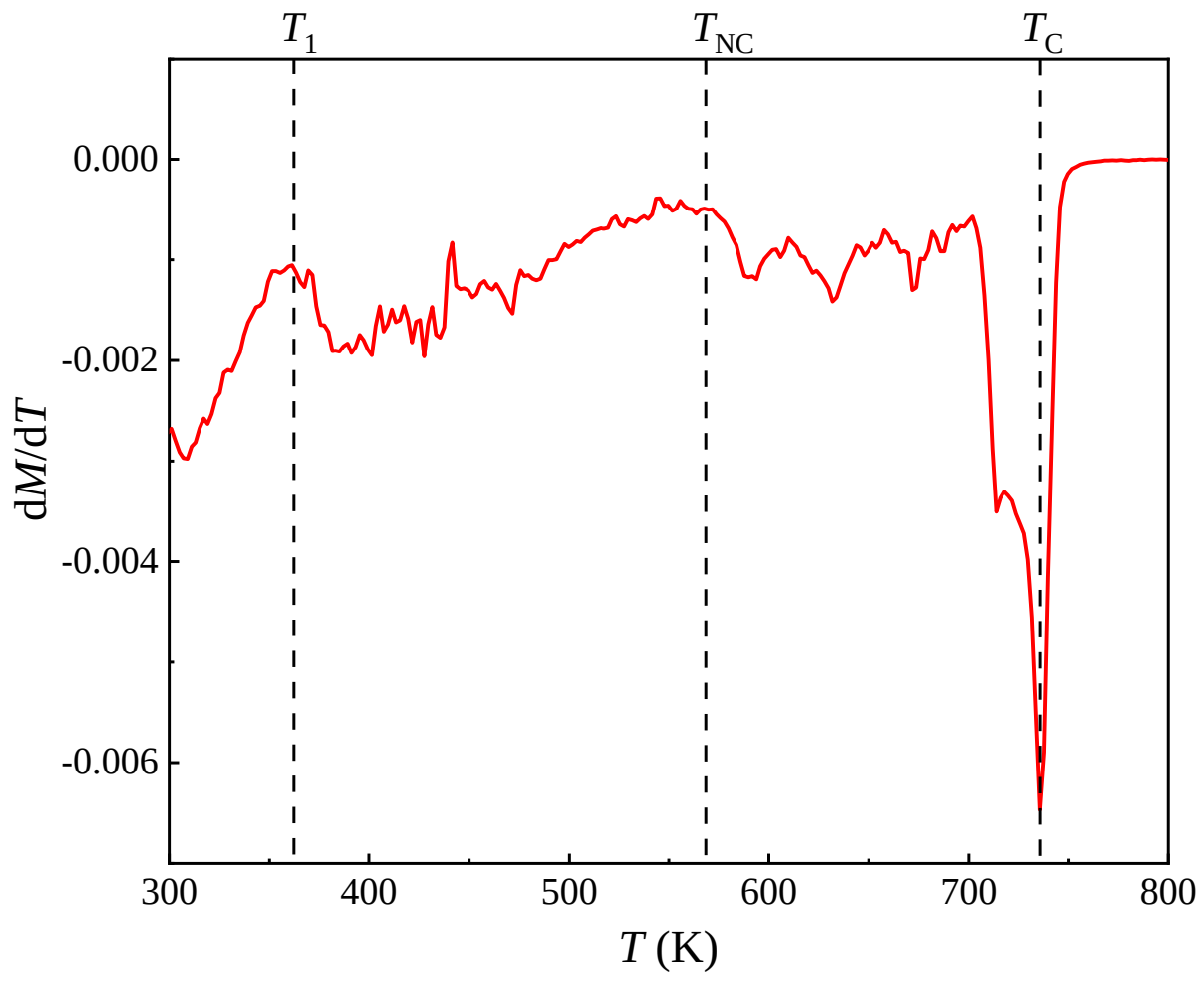

Figure S3. The first derivation of *M* vs. *T* curve.

***H* dependence of the relative dielectric constant $\varepsilon_r$:** During the measurement process, $H \perp E$ was maintained, with a measurement frequency of 1 MHz. Measurements were conducted within a magnetic field range of -50 kOe to 50 kOe.

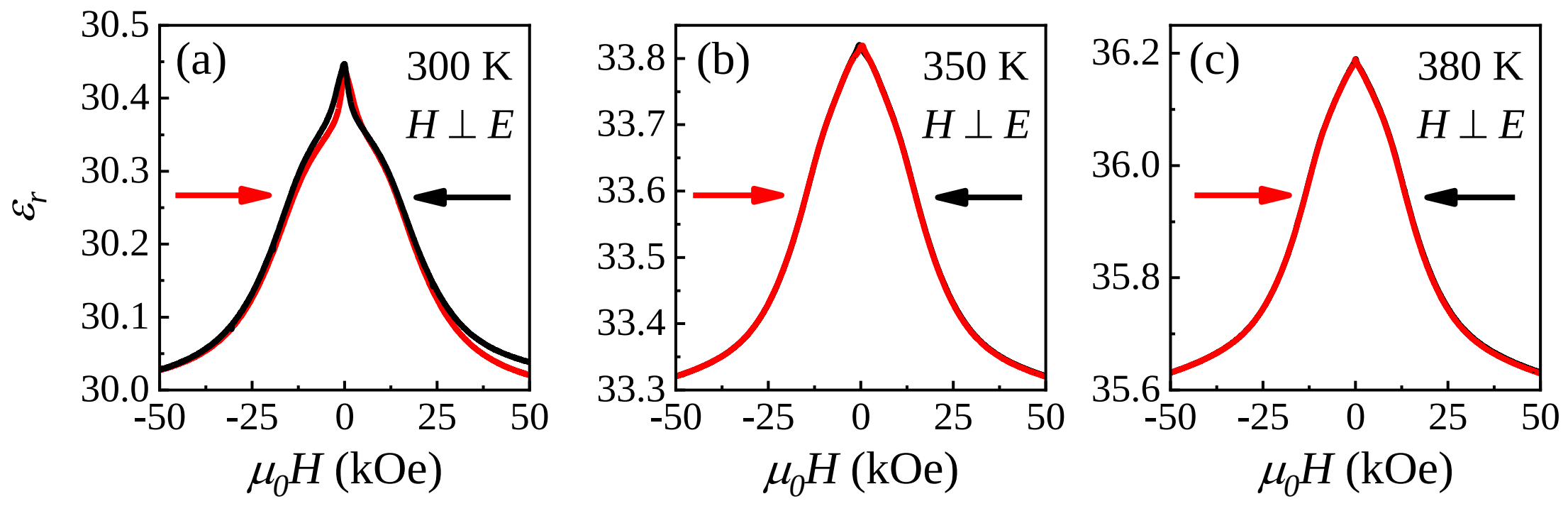

Figure S4. The curve of relative permittivity versus magnetic field for the sample at 300 K, 350 K, and 380 K.

Figure S4 displays the magnetic dielectric data of the sample at high temperatures. As the magnetic field increases, the sample transitions from the ferroelectric phase to the paraelectric phase. The strong magnetodielectric peak at zero field of the sample persist up to 380 K, indicating that the $BaSrMg_{0.28}Co_{1.72}Fe_{10}Al_2O_{22}$ still exhibits ferroelectricity and noncollinear phase above $T_1$.

***H* reversal of the electrical polarization *P*:** The polarization (*P*) was obtained by measuring the displacement current using the electrometer during the scanning of the magnetic field and then integrating the obtained data. Before the ME current measurements, we applied an electric field of 43.5 kV/m and a magnetic field of 5 T ($E \perp H$) to the sample at a fixed temperature. After reducing the magnetic field *H* to 0.5 T, remove the electric field *E* and then short-circuit the sample for 30 minutes, allowing it to discharge. Finally, the ME current was measured by scanning from -5 kOe to 5 kOe at a rate of 50 Oe/s.

Table S1. Summary of extracted and the descriptor predicted values of $T_{NC}$ of BaMgFeO-based Y-type hexaferrites.

| Number | Chemical composition | Extracted $T_{NC}$ (K) | References | Calculated $T_{NC}$ (K) |
|---|---|---|---|---|
| 1 | $Ba_2Ni_2Fe_{12}O_{22}$ | 0 | [1] | 11.56 |
| 2 | $Ba_{1.75}Sr_{0.25}Ni_2Fe_{12}O_{22}$ | 0 | [1] | 64.59625 |
| 3 | $Ba_{1.5}Sr_{0.5}Ni_2Fe_{12}O_{22}$ | 0 | [1] | 117.6325 |
| 4 | $Ba_{1.4}Sr_{0.6}Ni_2Fe_{12}O_{22}$ | 0 | [1] | 138.847 |
| 5 | $Ba_{1.25}Sr_{0.75}Ni_2Fe_{12}O_{22}$ | 88 | [1] | 170.66875 |
| 6 | $BaSrNi_2Fe_{12}O_{22}$ | 168 | [1] | 223.705 |
| 7 | $Ba_{0.75}Sr_{1.25}Ni_2Fe_{12}O_{22}$ | 290 | [1] | 276.74125 |
| 8 | $Ba_{0.5}Sr_{1.5}Ni_2Fe_{12}O_{22}$ | 300 | [1] | 329.7775 |
| 9 | $Ba_2Ni_{1.6}Zn_{0.4}Fe_{12}O_{22}$ | 0 | [2] | 11.56 |
| 10 | $Ba_2Ni_{1.2}Zn_{0.8}Fe_{12}O_{22}$ | 0 | [2] | 11.56 |
| 11 | $Ba_2Ni_{0.8}Zn_{1.2}Fe_{12}O_{22}$ | 0 | [2] | 11.56 |
| 12 | $Ba_2Ni_{0.4}Zn_{1.6}Fe_{12}O_{22}$ | 120 | [2] | 11.56 |
| 13 | $Ba_2Zn_2Fe_{12}O_{22}$ | 0 | [2] | 11.56 |
| 14 | $Ba_{0.5}Sr_{1.5}Zn_2Fe_{12}O_{22}$ | 317 | [3] | 329.7775 |
| 15 | $Ba_{0.5}Sr_{1.5}Zn_2Fe_{11.95}Ni_{0.05}O_{22}$ | 322 | [3] | 326.24175 |
| 16 | $Ba_{0.5}Sr_{1.5}Zn_2Fe_{11.9}Ni_{0.1}O_{22}$ | 324 | [3] | 322.706 |
| 17 | $Ba_{0.5}Sr_{1.5}Zn_2Fe_{11.8}Ni_{0.2}O_{22}$ | 329 | [3] | 315.6345 |
| 18 | $Ba_{0.5}Sr_{1.5}Zn_2Fe_{11.65}Ni_{0.35}O_{22}$ | 326 | [3] | 305.02725 |
| 19 | $Ba_{0.5}Sr_{1.5}Zn_2Fe_{11.5}Ni_{0.5}O_{22}$ | 321 | [3] | 294.42 |
| 20 | $Ba_{0.5}Sr_{1.5}ZnNiFe_{12}O_{22}$ | 282 | [4] | 329.7775 |
| 21 | $Ba_{0.5}Sr_{1.5}Zn_{0.5}Ni_{1.5}Fe_{12}O_{22}$ | 284 | [4] | 329.7775 |
| 22 | $Ba_{0.5}Sr_{1.5}Zn_{1.6}Mg_{0.4}Fe_{12}O_{22}$ | 346 | [5] | 358.0635 |
| 23 | $Ba_{0.5}Sr_{1.5}Zn_{1.2}Mg_{0.8}Fe_{12}O_{22}$ | 368 | [5] | 386.3495 |
| 24 | $Ba_2(Zn_{0.8}Mg_{0.2})_2Fe_{12}O_{22}$ | 0 | [6] | 39.846 |
| 25 | $Ba_2(Zn_{0.6}Mg_{0.4})_2Fe_{12}O_{22}$ | 0 | [6] | 68.132 |
| 26 | $Ba_2(Zn_{0.52}Mg_{0.48})_2Fe_{12}O_{22}$ | 45 | [6] | 79.4464 |
| 27 | $Ba_2(Zn_{0.437}Mg_{0.563})_2Fe_{12}O_{22}$ | 80 | [6] | 91.18509 |
| 28 | $Ba_2(Zn_{0.368}Mg_{0.632})_2Fe_{12}O_{22}$ | 103 | [6] | 100.94376 |
| 29 | $Ba_2Mg_{1.4}Zn_{0.6}Fe_{12}O_{22}$ | 117 | [6] | 110.561 |
| 30 | $Ba_2(Zn_{0.209}Mg_{0.791})_2Fe_{12}O_{22}$ | 150 | [6] | 123.43113 |
| 31 | $Ba_2Mg_{1.6}Zn_{0.4}Fe_{12}O_{22}$ | 145 | [6] | 124.704 |
| 32 | $Ba_2(Zn_{0.109}Mg_{0.891})_2Fe_{12}O_{22}$ | 175 | [6] | 137.57413 |
| 33 | $Ba_2Mg_{1.8}Zn_{0.2}Fe_{12}O_{22}$ | 170 | [6] | 138.847 |
| 34 | $Ba_2Mg_2Fe_{12}O_{22}$ | 195 | [6] | 152.99 |
| 35 | $Ba_2Co_2Fe_{12}O_{22}$ | 225 | [7] | 152.99 |
| 36 | $Ba_2Co_{1.8}Zn_{0.2}Fe_{12}O_{22}$ | 200 | [7] | 138.847 |
| 37 | $Ba_2Co_{1.4}Zn_{0.6}Fe_{12}O_{22}$ | 160 | [7] | 110.561 |
| 38 | $Ba_2CoZnFe_{12}O_{22}$ | 119 | [7] | 82.275 |
| 39 | $Ba_2Co_{0.5}Zn_{1.5}Fe_{12}O_{22}$ | 60 | [7] | 46.9175 |

| Number | Chemical composition | Extracted $T_{NC}$ (K) | References | Calculated $T_{NC}$ (K) |
|---|---|---|---|---|
| 40 | $Ba_{1.5}Sr_{0.5}Co_{1.6}Zn_{0.4}Fe_{12}O_{22}$ | 192 | [8] | 230.7765 |
| 41 | $Ba_2Zn_{0.4}Co_{1.6}Fe_{12}O_{22}$ | 183 | [8] | 124.704 |
| 42 | $Ba_{0.5}Sr_{1.5}Zn_2(Fe_{0.96}Al_{0.04})_{12}O_{22}$ | 280 | [9] | 295.8343 |
| 43 | $Ba_{0.5}Sr_{1.5}Zn_2(Fe_{0.92}Al_{0.08})_{12}O_{22}$ | 300 | [9] | 261.8911 |
| 44 | $Ba_{0.5}Sr_{1.5}Zn_2(Fe_{0.88}Al_{0.12})_{12}O_{22}$ | 290 | [9] | 227.9479 |
| 45 | $Ba_{0.7}Sr_{1.3}Zn_2(Al_{0.02}Fe_{0.98})_{12}O_{22}$ | 293 | [10] | 270.3769 |
| 46 | $Ba_{0.7}Sr_{1.3}Zn_2(Al_{0.04}Fe_{0.96})_{12}O_{22}$ | 264 | [10] | 253.4053 |
| 47 | $Ba_{0.7}Sr_{1.3}Zn_2(Al_{0.06}Fe_{0.94})_{12}O_{22}$ | 220 | [10] | 236.4337 |
| 48 | $Ba_{0.7}Sr_{1.3}Zn_2(Al_{0.08}Fe_{0.92})_{12}O_{22}$ | 180 | [10] | 219.4621 |
| 49 | $Ba_{0.7}Sr_{1.3}Zn_2(Al_{0.1}Fe_{0.9})_{12}O_{22}$ | 155 | [10] | 202.4905 |
| 50 | $Ba_{0.7}Sr_{1.3}Zn_2(Al_{0.12}Fe_{0.88})_{12}O_{22}$ | 106 | [10] | 205.3191 |
| 51 | $Ba_{1.1}Sr_{0.9}Co_2Fe_{11}AlO_{22}$ | 360 | [11] | 414.6355 |
| 52 | $Ba_{0.5}Sr_{1.5}Co_2Fe_{11}AlO_{22}$ | 430 | [12] | 400.4925 |
| 53 | $Ba_{0.8}Sr_{1.2}Co_2Al_{0.9}Fe_{11.1}O_{22}$ | 400 | [13] | 414.6355 |
| 54 | $BaSrCo_2Al_{0.9}Fe_{11.1}O_{22}$ | 420 | [13] | 428.7785 |
| 55 | $Ba_{1.2}Sr_{0.8}Co_2Al_{0.9}Fe_{11.1}O_{22}$ | 410 | [13] | 386.3495 |
| 56 | $Ba_2Mg_2(Fe_{0.96}Mn_{0.04})_{12}O_{22}$ | 200 | [14] | 186.9332 |
| 57 | $Ba_2Mg_2(Fe_{0.92}Mn_{0.08})_{12}O_{22}$ | 205 | [14] | 220.8764 |
| 58 | $Ba_2Mg_2(Fe_{0.88}Mn_{0.12})_{12}O_{22}$ | 208 | [14] | 254.8196 |
| 59 | $Ba_{1.5}Sr_{0.5}Mg_2Fe_{12}O_{22}$ | 290 | [15] | 259.0625 |
| 60 | $BaSrMg_2Fe_{12}O_{22}$ | 355 | [15] | 365.135 |
| 61 | $Ba_{0.5}Sr_{1.5}Mg_2Fe_{12}O_{22}$ | 390 | [15] | 329.7775 |
| 62 | $Ba_{0.4}Sr_{1.6}Mg_2Fe_{12}O_{22}$ | 370 | [16] | 322.706 |
| 63 | $BaSrZnMgFe_{12}O_{22}$ | 310 | [17] | 294.42 |
| 64 | $Ba_{0.5}Sr_{1.5}Co_2Fe_{12}O_{22}$ | 310 | [18] | 329.7775 |
| 65 | $Ba_{0.3}Sr_{1.7}Co_2Fe_{12}O_{22}$ | 340 | [19] | 315.6345 |
| 66 | $Ba_{0.5}Sr_{1.5}Zn_2Fe_{11.4}Cr_{0.6}O_{22}$ | 263 | [20] | 287.3485 |
| 67 | $(Ba_{0.729}Sr_{0.271})_2Zn_2Fe_{12}O_{22}$ | 40 | [21] | 126.54259 |
| 68 | $(Ba_{0.595}Sr_{0.405})_2Zn_2Fe_{12}O_{22}$ | 120 | [21] | 183.39745 |
| 69 | $(Ba_{0.525}Sr_{0.475})_2Zn_2Fe_{12}O_{22}$ | 350 | [21] | 213.09775 |
| 70 | $(Ba_{0.414}Sr_{0.586})_2Zn_2Fe_{12}O_{22}$ | 350 | [21] | 260.19394 |
| 71 | $Ba_{0.7}Sr_{1.3}Zn_2Fe_{12}O_{22}$ | 310 | [10] | 287.3485 |
| 72 | $Ba_{0.6}Sr_{1.4}Zn_2Fe_{12}O_{22}$ | 331 | [22] | 308.563 |
| 73 | $(Ba_{0.252}Sr_{0.748})_2Zn_2Fe_{12}O_{22}$ | 315 | [21] | 328.92892 |
| 74 | $(Ba_{0.193}Sr_{0.807})_2Zn_2Fe_{12}O_{22}$ | 280 | [21] | 353.96203 |
| 75 | $Ba_2Mg_{0.5}Co_{1.5}Fe_{12}O_{22}$ | 206 | [23] | 152.99 |
| 76 | $BaSrCo_2Fe_{11}AlO_{22}$ | 426 | [24] | 435.85 |
| 77 | $BaSrZn_{0.4}Co_{1.6}AlFe_{11}O_{22}$ | 400 | [24] | 407.564 |
| 78 | $BaSrZn_{0.8}Co_{1.2}AlFe_{11}O_{22}$ | 364 | [24] | 379.278 |
| 79 | $BaSrZn_{1.2}Co_{0.8}AlFe_{11}O_{22}$ | 334 | [24] | 322.706 |
| 80 | $BaSrZn_{1.6}Co_{0.4}AlFe_{11}O_{22}$ | 291 | [24] | 237.848 |
| 81 | $BaSrCoZnFe_{11}AlO_{22}$ | 383 | [25] | 365.135 |
| 82 | $BaSrZn_2AlFe_{11}O_{22}$ | 241 | [24] | 152.99 |
| Number | Chemical composition | Extracted $T_{NC}$ (K) | References | Calculated $T_{NC}$ (K) |
| 83 | $Ba_{1.3}Sr_{0.7}Co_{0.9}Zn_{1.1}Fe_{10.8}Al_{1.2}O_{22}$ | 359 | [26] | 308.563 |

Table S2. Summary of the maximum magnetoelectric coefficients α and the reported $T_{NC}$ of BaMgFeO-based Y-type hexaferrites in polycrystalline (Poly.) or single crystal (Single) form.

| Number | Chemical composition | $T_{NC}$ (K) | $\alpha$ (ps/m) | References |
|---|---|---|---|---|
| 1 | $Ba_{1.1}Sr_{0.9}Co_2Fe_{11}AlO_{22}$ (Poly.) | 360 | Converse 1200 @300 K | [11] |
| 2 | $BaSrZnMgFe_{12}O_{22}$ (Poly.) | 310 | 17.3 @20 K | [17] |
| 3 | $Ba_{1.3}Sr_{0.7}Co_{0.9}Zn_{1.1}Fe_{10.8}Al_{1.2}O_{22}$ (Single) | 359 | 1065 @200 K | [26] |
| 4 | $Ba_{0.4}Sr_{1.6}Mg_2Fe_{12}O_{22}$ (Single) | 370 | 33000 @10 K | [27] |
| 5 | $BaSrCoZnFe_{11}AlO_{22}$ (Poly.) | - | 3000 @200 K | [28] |
| 6 | $BaSrCo_2Fe_{11}AlO_{22}$ (Oriented Poly.) | - | 15000 @200 K | [29] |
| 7 | $Ba_{0.5}Sr_{1.5}Zn_2(Fe_{0.92}Al_{0.08})_{12}O_{22}$ (Single) | - | 20000 @15 K | [30] |
| 8 | $Ba_{0.5}Sr_{1.5}Zn_2Fe_{12}O_{22}$ (Single) | - | 1300 @30 K | [30] |
| 9 | $Ba_{0.5}Sr_{1.5}Co_2Fe_{11}AlO_{22}$ (Poly.) | - | Converse 101 @300 K | [31] |
| 10 | $BaSrMg_{0.28}Co_{1.72}Fe_{10}Al_2O_{22}$ (Poly.) | 568 | 44.4 @100 K | This work |

## Supplementary References